\documentclass[a4paper]{article}

\usepackage{INTERSPEECH2022}

\usepackage{amsmath,amsfonts,amssymb,graphicx,bm,enumerate,pifont,adjustbox,diagbox}
\usepackage[outline]{contour}
\usepackage[linesnumbered,ruled,vlined]{algorithm2e}
\usepackage{array}
\usepackage{multirow}

\makeatletter
\newcommand{\thickhline}{%
    \noalign {\ifnum 0=`}\fi \hrule height 1pt
    \futurelet \reserved@a \@xhline
}
\newcolumntype{"}{@{\hskip\tabcolsep\vrule width 1pt\hskip\tabcolsep}}
\makeatother

\title{Self-regularised Minimum Latency Training for Streaming Transformer-based Speech Recognition}
\name{Mohan Li, Rama Doddipatla and C\u{a}t\u{a}lin Zoril\u{a}}
\address{Cambridge Research Laboratory, Toshiba Europe Ltd, Cambridge, UK}
\email{\{mohan.li,rama.doddipatla,catalin.zorila\}@crl.toshiba.co.uk}

\begin{document}

\maketitle
\begin{abstract}

This paper proposes a self-regularised minimum latency training (SR-MLT) method for streaming Transformer-based automatic speech recognition (ASR) systems. In previous works, latency was optimised by truncating the online attention weights based on the hard alignments obtained from conventional ASR models, without taking into account the potential loss of ASR accuracy. On the contrary, here we present a strategy to obtain the alignments as a part of the model training without external supervision. The alignments produced by the proposed method are dynamically regularised on the training data, such that the latency reduction does not result in the loss of ASR accuracy. SR-MLT is applied as a fine-tuning step on the pre-trained Transformer models that are based on either monotonic chunkwise attention (MoChA) or cumulative attention (CA) algorithms for online decoding. ASR experiments on the AIShell-1 and Librispeech datasets show that when applied on a decent pre-trained MoChA or CA baseline model, SR-MLT can effectively reduce the latency with the relative gains ranging from 11.8\% to 39.5\%. Furthermore, we also demonstrate that under certain accuracy levels, the models trained with SR-MLT can achieve lower latency when compared to those supervised using external hard alignments.

\end{abstract}
\noindent\textbf{Index Terms}: streaming Transformer-based ASR, self-regularised minimum latency training

\section{Introduction}

The Transformer architecture \cite{vaswani2017attention}, as an epitome of the attention-based encoder-decoder framework \cite{chan2016listen, chorowski2015attention}, has become one of the dominant end-to-end (E2E) ASR techniques. Compared with other modelling strategies such as connectionist temporal classification (CTC) \cite{graves2006connectionist, graves2014towards} and recurrent neural network (RNN) transducer \cite{graves2012sequence}, Transformer sustains severe latency issues at inference time, since it requires access to the full speech utterance for decoding. To facilitate streaming ASR, a number of online attention mechanisms have been proposed in literature, including monotonic chunkwise attention (MoChA) \cite{chiu2017monotonic, tsunoo2019towards, inaguma2020enhancing}, CTC-triggered attention \cite{moritz2020streaming}, decoder-end adaptive computation steps (DACS) based algorithms \cite{li2021transformer, li2021head}, as well as the recent cumulative attention (CA) \cite{li2022transformer}. Although the aforementioned algorithms enable Transformer to emit ASR outputs in real time, the emission can be delayed from the actual acoustic boundary, because the system always tends to involve some future information to enhance the prediction confidence. 

There have been attempts to reduce the latency level of attention-based online decoding, and the most popular method is minimum latency training (MLT) \cite{inaguma2020minimum}. MLT introduces the "ground-truth" token boundaries into the attention calculation, which can be obtained from the forced alignments produced by conventional hidden Markov model (HMM) based ASR models \cite{chiu2018state}. When training the E2E system, any attending probabilities beyond the boundary will be ignored, so as to attenuate the chance of triggering later than the real acoustic border.

Besides, the CTC posterior spikes corresponding to the non-blank labels are also regarded as effective attention boundaries used in MLT, given that they sparsely appear around the endpoint of acoustic events \cite{moritz2020streaming}. Similar to the HMM-based method, a Viterbi decoding can be performed with the CTC model to generate the alignments that locate the output tokens. As a result, the triggering points produced by the attention model are calibrated as close to the CTC spikes as possible \cite{miao2022low}.

In MLT, both HMM and CTC alignments impose arbitrary restrictions on the attention computation, where the effect of latency reduction is largely subject to the quality of the alignments themselves, which determine the upper bound of MLT performance. Furthermore, the sacrifice of ASR accuracy in exchange of lower latency is not well considered by MLT, making the trade-off easily biased to the latency side.

Deep reinforcement learning boosted head-synchronous DACS (DRL-HS-DACS) \cite{li2021improving}, on the contrary, saves acquiring prior knowledge by self-exploring dynamic attention boundaries through its own learning experience. During E2E training, the system keeps track of the halting position along with the ASR prediction correctness observed on the training data, based on which the halting agent learns to seek the earliest triggering point with sufficient acoustic information. As opposed to MLT methods, DRL-HS-DACS could not only breakthrough the lower limit of latency posed by external alignments, but also maintain a balance between the accuracy and latency levels.

Even if the halting positions produced by the DRL-HS-DACS model are generally optimised by the DRL algorithm, the absence of strong constraints (e.g. hard acoustic boundaries) upon the agent training can still lead to the failure of halting for certain decoding steps. Thus, other auxiliary halting strategies such as the maximum look-ahead steps are also adopted to help the system trigger ASR outputs in time.

To overcome the issues encountered by MLT and DRL-HS-DACS while take advantage of their strengths, we propose the self-regularised minimum latency training (SR-MLT) algorithm. On the one hand, hard acoustic boundaries are utilised to cut off the redundant attention weights that cause delayed triggering, eliminating the need of DRL. On the other hand, these boundaries are dynamically obtained from the course of E2E training itself, rather than from external resources like forced alignments, which notably cuts down the training complexity when compared with the MLT methods.

We applied SR-MLT on two streaming Transformer ASR systems that are based on MoChA \cite{chiu2017monotonic} and CA \cite{li2022transformer} algorithms respectively, in order to show it compatible to various online attention mechanisms. Our experiments demonstrate that the proposed method can achieve significant latency reduction against the baseline systems. including the vanilla MLT systems that leverage HMM produced acoustic boundaries.

\section{Streaming Transformer-based ASR}

\subsection{Model architecture}

The Transformer ASR system is composed of three parts, as front-end, encoder and decoder. The front-end is usually implemented by convolutional neural networks (CNNs), which enhance the feature extraction and conduct frame-rate reduction. A stack of identical layers construct the Transformer encoder, with each consisting of a self-attention module and a pointwise feed-forward network (FFN). The decoder takes a similar architecture, but having an additional cross-attention module at each layer, where the speech-to-text alignment is generated.

Transformer adopts the scaled dot-product attention mechanism to capture the dependency between sequence elements:
\begin{equation}
    \mathrm{Attention}(\mathbf{Q},\mathbf{K},\mathbf{V}) = \mathrm{softmax}(\frac{\mathbf{Q} \mathbf{K}^\mathrm{T}} {\sqrt{d_k}}) \mathbf{V}, 
    \label{eq: dotproductatt} 
\end{equation}
where $\mathbf{Q,K,V}$ denote the projected feature representations within either the encoder or decoder attention modules, given $d_k$ as the dimension of projection. Moreover, multiple attention heads are utilised by Transformer, aiming to obtain information from distinct feature spaces:
\begin{equation} 
    \mathrm{MultiHead}(\mathbf{Q},\mathbf{K},\mathbf{V}) = \mathrm{Concat}(\mathrm{head}_1,...,\mathrm{head}_H) \mathbf{W}^O, 
\end{equation}
\begin{equation} 
    \mathrm{and\;head}_h = \mathrm{Attention}(\mathbf{QW}^Q_h,\mathbf{KW}^K_h,\mathbf{VW}^V_h), 
\end{equation}
where $\mathbf{W}^{Q,K,V}_h$ and $\mathbf{W}^O$ denote the linear projection matrices of the individual heads and the output layer.

The standard Transformer system performs global attention on the entire input sequence, which precludes the streaming inference of ASR. At the encoder side, the solution can be segmenting the speech utterance into a steam of chunks with fixed \cite{tsunoo2019transformer} or adaptive \cite{li2019end, dong2020cif} size, so that the attention computation is restricted to local contexts. As for the decoder side is concerned, a few online decoding algorithms have been proposed, and here we present two of them as follows.

\subsection{Monotonic chunkwise attention (MoChA)}

Rather than working on the full speech utterance, MoChA first estimates the probability of stopping the decoding at each frame, which is monotonically computed along the timestep $j$, as:
\begin{equation}
    p_{i,j} = \mathrm{Sigmoid} \left( e_{i,j} \right),
    \label{eq:att_prob}
\end{equation}
\begin{equation}
    e_{i,j} = \frac {\bm{q}_i (\bm{k}_{i,j})^T} {\sqrt{d_k}},
    \label{eq:att_en}
\end{equation}
where $\bm{q}$ and $\bm{k}$ denote the decoder and encoder states seen at the decoding step $i$, respectively. The sigmoid function here serves as a substitute to the softmax function in eq. (\ref{eq: dotproductatt}), performing instant normalisation. To facilitate the system training, an expectation form of the triggering probability is given as:
\begin{equation}
    \alpha_{i,j} = p_{i,j} \left( (1-p_{i,j-1})\frac{\alpha_{i,j-1}}{p_{i,j-1}} + \alpha_{i-1,j} \right),
    \label{eq:alpha}
\end{equation}
which can be recursively calculated over the decoding steps.

As the name suggests, MoChA also conducts a second pass soft attention upon the chunks of encoder states, in order to add flexibility to the speech-to-text alignment:
\begin{equation}
    \beta_{i,j} = \left. \sum^{j+w-1}_{k=j} \left( \alpha_{i,k}\mathrm{exp}(u_{i,j}) \middle/ \sum^k_{l=k-w+1} {\mathrm{exp}(u_{i,l})} \right) \right.,
\end{equation}
given $w$ as the width of chunks, and $u_{i,j}$ the chunkwise attention energy that is computed as in eq. (\ref{eq:att_en}).

Finally, a context vector is generated as the weighted sum of the encoder states $\bm{v}$:
\begin{equation} 
    \bm{c}_i = \sum_{j=1}^T \beta_{i,j} \cdot \bm{v}_j, 
    \label{eq:mocha_cv}
\end{equation}
where $T$ denotes the length of speech.

During test time, the earliest timestep that has $p_{i,j} > 0.5$ will be chosen as the decoding endpoint, and the chunk ending at the corresponding $j$ is used to produce the context vector.

\subsection{Cumulative attention (CA)}

The CA algorithm has been proposed to mitigate the problem of MoChA that the decoding is halted only based on the acoustic feature of single frames. Instead, the halting decisions in CA are made upon all history encoder states.

Contrary to MoChA, CA works out the interim context vectors in the first place, following the autoregressive manner:
\begin{equation}
    \bm{c}_{i,j} = \bm{c}_{i, j-1} + a_{i,j} \cdot \bm{v}_{i,j},
    \label{eq:int_cv}
\end{equation}
where $a_{i,j}$ is computed in the same way as eq. (\ref{eq:att_prob}), but rather interpreted as the monotonic attention weight. It is understood that $\bm{c}_{i,j}$ contains all the relevant acoustics information accumulated at each timestep. Next, a trainable device called halting selector is introduced to take $\bm{c}_{i,j}$ as input, from which a halting probability is calculated:
\begin{equation}
    p_{i,j} = \mathrm{Sigmoid} (\mathrm{HaltSelect}(\bm{c}_{i,j}) + r + \epsilon),
    \label{eq:hp}
\end{equation}
where $r$ is a bias term and $\epsilon$ denotes the additive Gaussian noise applied during training. The halting selector is usually implemented as a simple one-layer deep neural network (DNN) to speed up the inference. Similar to MoChA, an expected halting probability $\alpha_{i,j}$ is computed through eq. (\ref{eq:alpha}) to enable the update of halting selector parameters.

As a result, the expected context vector is produced to cover all the alignment paths as:
\begin{equation}
    \bm{c}_i = \sum^T_{j=1} \alpha_{i,j} \cdot \bm{c}_{i,j}.
    \label{eq:cv}
\end{equation}

For inference, the interim context vectors are dynamically updated along the timesteps as in training, and likewise the first one with $p_{i,j} > 0.5$ will be directly picked up as the output of the attention module, which concludes the decoding.

\section{Proposed self-regularised minimum latency training}

The goal of SR-MLT is to achieve the maximum latency reduction during inference at the minimum cost of ASR accuracy. Thus, we must carefully balance both metrics when designing the method. To accomplish this, SR-MLT is carried out in two stages: pre-training and fine-tuning.

\subsection{Pre-training}

As no external alignments are leveraged, the SR-MLT method has to obtain the appropriate acoustic boundaries completely on its own. We assume that the triggering points (halting positions) that correspond to accurate ASR predictions can be used as good initial boundaries for the later fine-tuning stage. Thus, in the course of pre-training, we train the online Transformer system as normal with MoChA or CA applied, and keep recording (a) the triggering points and (b) the ASR prediction correctness of each decoding step within the training set. Based on these factors, the accuracy and coverage ratio (which reflects latency) of the data are worked out:
\begin{equation}
    acc = \frac{\text{number of correctly predicted tokens}}{\text{total number of tokens}},
    \label{eq:acc}
\end{equation}
\begin{equation}
     cov = \frac{\text{triggering point index}}{\text{total length of utterance}}.
    \label{eq:cov}
\end{equation}
The recorded factors will be updated as soon as a higher accuracy is seen at the current data iteration when compared with the record, regardless of the coverage ratio variation, or otherwise remain unchanged to the previous figures.

Specifically, compared to \cite{li2021improving} where the metrics in eq. (\ref{eq:acc}) and (\ref{eq:cov}) are calculated with respect to individual tokens, SR-MLT calculates the metrics upon the whole mini-batch of training data. We believe that the statistics measured on larger granularity, such as mini-batch, can provide a more reliable estimate toward the system performance, so as to help make better update decisions during training. A comparison of ASR performance using different levels of granularity including token, utterance and mini-batch will be given in section 4.3. 

Note that the number of pre-training epochs can be neither too small, where the ASR model hasn't matured to reach an adequate accuracy level and provide stable triggering points, nor too large such that the model is over-fitted to the training data. In our experiments, the models are pre-trained up to around 60\% of the full training time. In the end, the pre-training stage evaluates the upper bound of the ASR accuracy that is achievable by the streaming system, which is exactly what we aim to hold to when reducing the latency at the next stage.

\begin{table}[t]
\centering
\caption{Update policy of SR-MLT, with regard to the prediction correctness and triggering point on training set.}
\begin{tabular}{l|lll}
\thickhline
\backslashbox{acc}{cov} & \makebox[3em]{equal} & \makebox[3em]{up} & \makebox[3em]{down} \\ \hline
\multicolumn{1}{l|}{equal}   & \multicolumn{1}{c}{False}   & \multicolumn{1}{c}{False}   & \multicolumn{1}{c}{True} \\
\multicolumn{1}{l|}{up}   & \multicolumn{1}{c}{True}   & \multicolumn{1}{c}{True}   & \multicolumn{1}{c}{True} \\
\multicolumn{1}{l|}{down} & \multicolumn{1}{c}{False} & \multicolumn{1}{c}{False}    & \multicolumn{1}{c}{False} \\
\thickhline
\end{tabular}
\label{tab:update}
\vspace{-4mm}
\end{table}

\subsection{Fine-tuning}

At the fine-tuning stage, the initial boundaries obtained from the pre-training are applied to restricting the attention computation, similar to \cite{inaguma2020minimum}:
\begin{equation}
    \alpha_{i,j} = 
        \begin{cases}
            p_{i,j} \left( (1-p_{i,j-1})\frac{\alpha_{i,j-1}}{p_{i,j-1}} + \alpha_{i-1,j} \right), & j \leq b_i + \delta \\
            0, & \text{otherwise}
        \end{cases}
    \label{eq:mlt}
\end{equation}
where $b_i$ is the triggering point in the record, and $\delta$ denotes an offset appending to the acoustic boundary which allows some extra frames to be included to the attention, easing the restriction imposed by hard alignments.

Meanwhile, we continue to record the prediction correctness and triggering point on the training set, but the update of these two factors follows the policy presented in Table. \ref{tab:update}. It can be seen that the accuracy still dictates the update decision. The records are made up-to-date whenever the accuracy goes up on the same mini-batch of data, or left untouched if it declines. As for the equal accuracy is concerned, the update is only performed when the coverage ratio shows a downward tendency, which indicates the reduction of latency label.

Once being updated, The new triggering point will be immediately applied to eq. (\ref{eq:mlt}) as $b_i$ for the following training epochs. By doing this, the acoustic boundaries could be gradually pushed to the lower limit as the training proceeds. However, in normal MLT, the streaming system may have rare opportunities to produce triggering points that are substantially earlier than the arbitrary HMM alignments. 

At the same time, SR-MLT doesn't lose the potential of achieving equally good ASR accuracy as before the fine-tuning, because the inappropriate triggering points that lead to decayed accuracy are precluded into becoming the valid acoustic boundaries. Hence, we could say that the ASR performance is self-regularised through such a balancing strategy.

\section{Experiments}

\subsection{Experimental setup}

The proposed method has been evaluated on two open-sourced datasets: AIShell-1 \cite{bu2017aishell} and Librispeech \cite{panayotov2015librispeech}. We adopt precisely the same data preparation, model architecture, and external language model as in \cite{li2022transformer} for fair comparisons. The number of pre-training and fine-tuning epochs, as well as the offset $\delta$ (after frame-rate reduction by front-end) are set to \{30, 20, 2\} for AIShell-1, and \{70, 50, 8\} for Librispeech. When recording the triggering point for each decoding step, since each timestep is assigned with a triggering (halting) probability $\alpha_{i,j}$ (see eq. (\ref{eq:alpha})) rather than a binary decision. we put the timestep with the largest $\alpha_{i,j}$ as the one into practice. Besides, to enhance the update rate in the fine-tuning stage, any variations of accuracy and coverage ratio that are less than 0.1\% are deemed as 'equal', given that the exact equality of the metrics can be very seldom seen on the larger mini-batch of speech utterances.

\begin{table}[t]
\centering
\caption{Character error rates (CERs \%) on AIShell-1.}
\begin{tabular}{llll}
\hline \thickhline
\multicolumn{1}{l}{Model}  &&  \multicolumn{1}{c}{dev} & \multicolumn{1}{c}{test} \\ \thickhline
\multicolumn{1}{l}{Offline \cite{karita2019comparative}}    & & \multicolumn{1}{c}{-}      & \multicolumn{1}{c}{6.7}       \\ \hline 
\multicolumn{1}{l}{MMA-MoChA \cite{inaguma2020enhancing}}      & & \multicolumn{1}{c}{-}      & \multicolumn{1}{c}{7.5}       \\
\multicolumn{1}{l}{BS-DEC \cite{tsunoo2021streaming}}    && \multicolumn{1}{c}{6.4}      & \multicolumn{1}{c}{7.3} \\
\multicolumn{1}{l}{MoChA \cite{li2022transformer}}  && \multicolumn{1}{c}{6.4}  & \multicolumn{1}{c}{7.2}\\
\multicolumn{1}{l}{\bf{SR-MLT MoChA}}  && \multicolumn{1}{c}{\bf{6.3}}  & \multicolumn{1}{c}{\bf{7.2}}\\
\multicolumn{1}{l}{CA \cite{li2022transformer}}  && \multicolumn{1}{c}{6.3}  & \multicolumn{1}{c}{7.0}\\ 
\multicolumn{1}{l}{\bf{SR-MLT CA}}  && \multicolumn{1}{c}{\bf{6.4}}  & \multicolumn{1}{c}{\bf{7.1}}\\ 
\hline \thickhline
\label{tab:aishell}
\end{tabular}

\centering
\caption{Word error rates (WERs \%) on Librispeech.}
\begin{tabular}{lcccc}
\hline \thickhline
\multirow{2}{*}{Model}      & \multicolumn{2}{c}{dev}   & \multicolumn{2}{c}{test}     \\ \cline{2-5} 
                        & \multicolumn{1}{c}{clean}    & \multicolumn{1}{c}{other} & \multicolumn{1}{c}{clean}    & \multicolumn{1}{c}{other} \\ \thickhline
\multicolumn{1}{l}{Offline \cite{li2022transformer}}  & \multicolumn{1}{c}{2.4}  & \multicolumn{1}{c}{6.0} 
                                        & \multicolumn{1}{c}{2.6}  & \multicolumn{1}{c}{6.1}    \\ \hline
\multicolumn{1}{l}{Triggered attention \cite{moritz2020streaming}}    & \multicolumn{1}{c}{-}  & \multicolumn{1}{c}{-} 
                                        & \multicolumn{1}{c}{2.8}  & \multicolumn{1}{c}{7.2} \\
\multicolumn{1}{l}{BS-DEC \cite{tsunoo2021streaming}} & \multicolumn{1}{c}{2.5} & \multicolumn{1}{c}{6.8} 
                                                                  & \multicolumn{1}{c}{2.7}  & \multicolumn{1}{c}{7.1} \\
\multicolumn{1}{l}{MoChA \cite{li2022transformer}}  & \multicolumn{1}{c}{2.8}  & \multicolumn{1}{c}{7.1}
                           & \multicolumn{1}{c}{3.1}  & \multicolumn{1}{c}{7.4} \\
\multicolumn{1}{l}{\bf{SR-MLT MoChA}}  & \multicolumn{1}{c}{\bf{2.8}}  & \multicolumn{1}{c}{\bf{7.4}} 
                                       & \multicolumn{1}{c}{\bf{3.1}}  & \multicolumn{1}{c}{\bf{7.5}} \\
\multicolumn{1}{l}{CA \cite{li2022transformer}}  & \multicolumn{1}{c}{2.5}  & \multicolumn{1}{c}{6.7}
                        & \multicolumn{1}{c}{2.7}  & \multicolumn{1}{c}{6.8} \\
\multicolumn{1}{l}{\bf{SR-MLT CA}}  & \multicolumn{1}{c}{\bf{2.6}}  & \multicolumn{1}{c}{\bf{6.8}} 
                                    & \multicolumn{1}{c}{\bf{2.7}}  & \multicolumn{1}{c}{\bf{7.1}} \\
\hline \thickhline
\label{tab:librispeech}
\end{tabular}
\vspace{-6mm}
\end{table}

\subsection{Experimental results}

Table. \ref{tab:aishell} and \ref{tab:librispeech} present the character-error-rate (CER) and the word-error-rate (WER) achieved by the proposed method on AIShell-1 and Librispeech datasets, respectively. From the results, we could find that SR-MLT has successfully maintained the accuracy level for the streaming ASR systems. On AIShell-1, the CERs of both SR-MLT MoChA and SR-MLT CA models have been preserved very closely to the baseline numbers, As for Librispeech is concerned, even if the accuracy of SR-MLT based models drops on the test-other set when compared to the vanilla MoChA and CA models, they are still kept in the similar range with the reference systems in literature.


\begin{table}[t]
\centering
\caption{Latency (frames) on AIShell-1.}
\begin{tabular}{llll}
\hline \thickhline
\multicolumn{1}{l}{Model}  &&  \multicolumn{1}{c}{dev} & \multicolumn{1}{c}{test} \\ \thickhline
\multicolumn{1}{l}{Offline \cite{li2022transformer}}    & & \multicolumn{1}{c}{232.8}      & \multicolumn{1}{c}{257.0}       \\ \hline
\multicolumn{1}{l}{MoChA \cite{li2022transformer}}  && \multicolumn{1}{c}{90.1}  & \multicolumn{1}{c}{92.5}\\
\multicolumn{1}{l}{\bf{SR-MLT MoChA}}  && \multicolumn{1}{c}{\bf{56.3}}  & \multicolumn{1}{c}{\bf{56.0}}\\
\multicolumn{1}{l}{CA \cite{li2022transformer}}  && \multicolumn{1}{c}{52.8}  & \multicolumn{1}{c}{51.8}\\ 
\multicolumn{1}{l}{\bf{SR-MLT CA}}  && \multicolumn{1}{c}{\bf{47.9}}  & \multicolumn{1}{c}{\bf{45.7}}\\ 
\hline \thickhline
\label{tab:aishell_lat}
\end{tabular}

\centering
\caption{Latency (frames) on Librispeech.}
\begin{tabular}{lcccc}
\hline \thickhline
\multirow{2}{*}{Model}      & \multicolumn{2}{c}{dev}   & \multicolumn{2}{c}{test}     \\ \cline{2-5} 
                        & \multicolumn{1}{c}{clean}    & \multicolumn{1}{c}{other} & \multicolumn{1}{c}{clean}    & \multicolumn{1}{c}{other} \\ \thickhline
\multicolumn{1}{l}{Offline \cite{li2022transformer}}  & \multicolumn{1}{c}{497.9}  & \multicolumn{1}{c}{440.7} 
                                        & \multicolumn{1}{c}{528.5}  & \multicolumn{1}{c}{463.8}    \\ \hline
\multicolumn{1}{l}{MoChA \cite{li2022transformer}}  & \multicolumn{1}{c}{295.2}  & \multicolumn{1}{c}{256.4}
                           & \multicolumn{1}{c}{303.1}  & \multicolumn{1}{c}{271.2} \\
\multicolumn{1}{l}{\bf{SR-MLT MoChA}}  & \multicolumn{1}{c}{\bf{295.0}}  & \multicolumn{1}{c}{\bf{256.5}} 
                                       & \multicolumn{1}{c}{\bf{303.3}}  & \multicolumn{1}{c}{\bf{270.5}} \\
\multicolumn{1}{l}{CA \cite{li2022transformer}}  & \multicolumn{1}{c}{68.1}  & \multicolumn{1}{c}{63.5}
                        & \multicolumn{1}{c}{68.5}  & \multicolumn{1}{c}{65.5} \\
\multicolumn{1}{l}{\bf{SR-MLT CA}}  & \multicolumn{1}{c}{\bf{51.3}}  & \multicolumn{1}{c}{\bf{50.0}} 
                                    & \multicolumn{1}{c}{\bf{50.6}}  & \multicolumn{1}{c}{\bf{50.6}} \\
\hline \thickhline
\label{tab:librispeech_lat}
\end{tabular}%
\vspace{-6mm}
\end{table}

The latency level during the inference of SR-MLT models are presented in Table. \ref{tab:aishell_lat} and \ref{tab:librispeech_lat} for the two datasets, in comparison with the vanilla MoChA and CA based counterparts. The evaluation of latency is adopted as the same in \cite{li2022transformer}, which measures the difference between the triggering point of the E2E model $\hat{b}^k_i$ and the ground-truth token boundary $b^k_i$:
\begin{equation}
    \triangle_{corpus} = \frac{1}{\sum^N_{k=1} |\bm{y}^k|} \sum^N_{k=1} \sum^{|\bm{y}^k|}_{i=1} (\hat{b}^k_i - b^k_i),
    \label{eq:latency}
\end{equation}
where $N$ denotes the total number of utterances in the dataset, and $|\bm{y}^k|$ is the number of tokens in each utterance. On the test set of AIShell-1, One can observe that SR-MLT has effectively reduced the latency of both streaming systems, with the relative gains of 39.5\% and 11.8\% for MoChA and CA respectively. The significant reduction of latency has also been seen by SR-MLT CA on Librispeech, where the relative gains of 26.1\% and 22.7\% are achieved on the test-clean and test-other sets. However, in contrast with the success on AIShell-1, SR-MLT does not bring obvious latency reductions to the MoChA model on Librispeech. This is due to that some attention heads of MoChA-based Transformer can behave very poorly \cite{li2022transformer}, where SR-MLT is not able to optimise their attention boundaries. Thus, we always need a decent baseline model to perform SR-MLT well at the stage of fine-tuning.

\subsection{Analysis}

\begin{figure}[t]
  \centering
  \includegraphics[width=\linewidth]{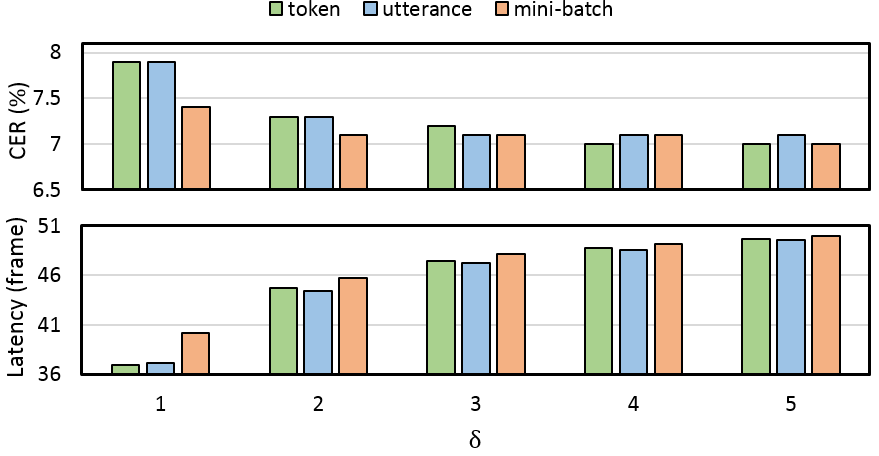}
  \caption{ASR performance of using different update granularity in SR-MLT CA on AIShell-1.}
  \label{fig:granularity}
\end{figure}

Figure. \ref{fig:granularity} illustrates the effect of using different granularity to calculate the accuracy and latency statistics on the training set as described in section 3.1. We experiment with SR-MLT CA on AIShell-1 at three levels, as token, utterance and mini-batch. Meanwhile, a range of offset values are manipulated to demonstrate the trade-off between the accuracy and latency metrics. From the figure, it is noticed that in general, using larger granularity could lead to improved accuracy and correspondingly compromised latency performance. Despite the harsh restrictions posed by smaller $\delta$ ($\delta=1$), the mini-batch setting is still able to contain the decay of accuracy under a reasonable level, as opposed to the token and utterance settings.

\begin{figure}[t]
  \centering
  \includegraphics[width=\linewidth]{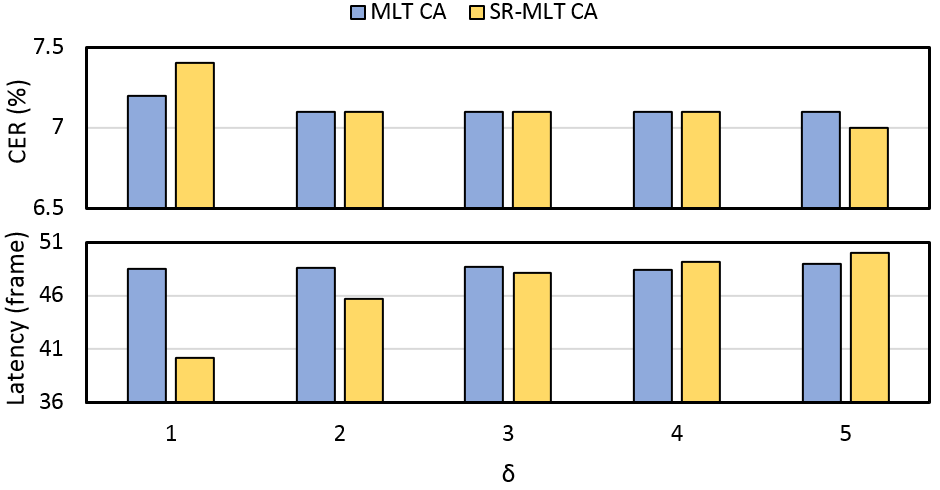}
  \caption{ASR performance of MLT CA and SR-MLT CA on AIShell-1.}
  \label{fig:comp}
  \vspace{-6mm}
\end{figure}

We further compare the performance of the proposed SR-MLT with the original MLT on top of the CA-base Transformer system. The results achieved on AIShell-1 are presented in Figure. \ref{fig:comp}. It can be observed that MLT is less sensitive to the change of $\delta$, with regard to both accuracy and latency levels. On the contrary, SR-MLT manages to acquire a better balance of the metrics as $\delta$ moderates. Specifically, when $\delta=2$ and 3, the SR-MLT model obtains lower latency with the same accuracy compared to the MLT model, showing the advantage of dynamic acoustic boundaries over the fixed HMM alignments.

\section{Conclusions}

In this work, we present a novel minimum latency training method, called SR-MLT, for streaming Transformer-based ASR. Compared to the original MLT method that leverages external forced alignments, SR-MLT produces dynamic acoustic boundaries fully from the model training, in order to restrict the frames that are involved to the online attention computation. The proposed method is conducted in two stages. At the pre-training stage, the optimal triggering points along with the ASR prediction correctness are recorded for all the training data. Then at the fine-tuning stage, the triggering points in the records are utilised as initial acoustic boundaries for the decoding steps. Meanwhile, as the training proceeds, the records are updated once reduced latency is observed on the training set with maintained accuracy. We applied SR-MLT onto two streaming ASR algorithms as MoChA and CA. The experimental results show that the proposed method can reduce the latency level for both MoChA and CA based systems, without degrading the ASR accuracy. Besides, we also show that SR-MLT could achieve a better balance between latency and accuracy when compared with normal MLT.

\bibliographystyle{IEEE}

\bibliography{refs}


\end{document}